\documentclass[fleqn,usenatbib]{mnras}

\usepackage{newtxtext,newtxmath}

\usepackage[T1]{fontenc}

\DeclareRobustCommand{\VAN}[3]{#2}
\let\VANthebibliography\thebibliography
\def\thebibliography{\DeclareRobustCommand{\VAN}[3]{##3}\VANthebibliography}

\usepackage{graphicx}	
\usepackage{amsmath}	

\usepackage{lipsum}
\usepackage[english]{babel}
\usepackage{color}
\usepackage[dvipsnames]{xcolor}

\usepackage{multirow}
\usepackage{cleveref}
\crefformat{section}{\S#2#1#3} 
\crefformat{subsection}{\S#2#1#3}
\crefformat{subsubsection}{\S#2#1#3}

\usepackage{romannum}

\title[47\,min DWD Binary]{An Eclipsing 47\,minute Double White Dwarf Binary at 400\,pc}
\author[J.\ Munday et al.]{James Munday,$^{1,2}$\thanks{Email: james.munday98@gmail.com}
P.-E. Tremblay,$^{1}$
J. J. Hermes,$^{3}$
Brad Barlow,$^{4}$
Ingrid Pelisoli,$^{1}$
T. R.\ Marsh,$^{1}$
\newauthor
Steven G. Parsons,$^{5}$
David Jones,$^{6,7,8}$
S. O. Kepler,$^{9}$
Alex Brown,$^{5}$
S. P.\ Littlefair,$^{5}$
R. Hegedus,$^{10}$
\newauthor
Andrzej Baran,$^{11,12}$
Elm\'{e} Breedt,$^{13}$
V. S.\ Dhillon,$^{5,6}$
Martin J. Dyer,$^{5}$
Matthew J. Green,$^{14}$
\newauthor
Mark R. Kennedy,$^{15,16}$
Paul Kerry,$^{5}$
Isaac D. Lopez,$^{17}$
Alejandra D. Romero,$^{9}$
Dave Sahman,$^{5}$
\newauthor
Hannah L. Worters$^{18}$\\
$^{1}$Department of Physics, Gibbet Hill Road, University of Warwick, Coventry CV4 7AL, United Kingdom\\
$^{2}$Isaac Newton Group of Telescopes, Apartado de Correos 368, E-38700 Santa Cruz de La Palma, Spain\\
$^{3}$Department of Astronomy \& Institute for Astrophysical Research, Boston University, 725 Commonwealth Ave., Boston, MA 02215, USA\\
$^{4}$Department of Physics and Astronomy, High Point University, High Point, NC, 27268, USA\\
$^{5}$Department of Physics and Astronomy, University of Sheffield, Sheffield, S3 7RH, UK\\
$^{6}$Instituto de Astrof\'isica de Canarias, E-38205 La Laguna, Tenerife, Spain\\
$^{7}$ Departamento de Astrof\'isica, Universidad de La Laguna, E-38206
La Laguna, Tenerife, Spain\\
$^{8}$ Nordic Optical Telescope, Rambla Jos\'e Ana Fern\'andez P\'erez
7, 38711, Bre\~na Baja, Spain\\
$^{9}$ Instituto de F\'isica, Universidade Federal do Rio Grande do Sul, Porto-Alegre, RS 91501-900, Brazil\\
$^{10}$Department of Physics and Astronomy, University of North Carolina at Chapel Hill, Chapel Hill, NC 27599, USA\\
$^{11}$ Astronomical Observatory, University of Warsaw, Al. Ujazdowskie 4, 00-478 Warszawa, Poland\\
$^{12}$ Department of Physics, Astronomy, and Materials Science, Missouri State University, Springfield, MO\,65897, USA\\
$^{13}$Institute of Astronomy, University of Cambridge, Madingley Road, Cambridge CB3 0HA, UK\\
$^{14}$Department of Astrophysics, School of Physics and Astronomy, Tel Aviv University, Tel Aviv 6997801, Israel\\
$^{15}$School of Physics, University College Cork, Cork, Ireland\\
$^{16}$Jodrell Bank Centre for Astrophysics, University of Manchester, Manchester M13 9PL, UK\\
$^{17}$Department of Physics and Astronomy, Iowa State University, Ames, IA 50011, USA\\
$^{18}$South African Astronomical Observatory, PO Box 9, Observatory 7935, Cape Town, South Africa
}

\date{Accepted 2023 July 31. Received 2023 July 27; in original form 2023 March 25}

\pubyear{2023}

\begin{document}
\label{firstpage}
\pagerange{\pageref{firstpage}--\pageref{lastpage}}
\maketitle

\begin{abstract}
We present the discovery of the eclipsing double white dwarf (WD) binary WDJ\,022558.21$-$692025.38 that has an orbital period of 47.19\,min. Following identification with the Transiting Exoplanet Survey Satellite, we obtained time-series ground based spectroscopy and high-speed multi-band ULTRACAM photometry which indicate a primary DA~WD of mass $0.40\pm0.04\,\text{M}_\odot$ and a $0.28\pm0.02\,\text{M}_\odot$ mass secondary WD, which is likely of type DA as well. The system becomes the third-closest eclipsing double WD binary discovered with a distance of approximately 400\,pc and will be a detectable source for upcoming gravitational wave detectors in the mHz frequency range. Its orbital decay will be measurable photometrically within 10\,yrs to a precision of better than 1\%. The fate of the binary is to merge in approximately 41\,Myr, likely forming a single, more massive WD.
\end{abstract}

\begin{keywords}
binaries: close --  individual: WDJ\,022558.21$-$692025.38 -- stars: white dwarfs -- gravitational waves
\end{keywords}



\section{Introduction}
Compact white dwarf (WD) binaries are of strong astrophysical interest as descendants of mass transfer phases, possibly the progenitors of exotic merger stars \citep[][]{Webbink1984DWDprogenitorsRCrB, Zhang2012FormationOfHeRichHotSubdwarfs,Cheng2019CoolingAnomalyOfHighMassWDs,Gvaramadze2019WDmergerproduct,Hollands2020UltraMassiveNature} and are one of the likely progenitors of type \Romannum{1}a/.\Romannum{1}a supernovae \citep[][]{Maoz2014Type1aProgenitors}. Among double WD (DWD) binary systems, the subset which exit mass transfer with periods less than $\sim10$ hours are destined to return into contact within a Hubble time, primarily as a consequence of gravitational wave radiation. There are hence many DWD binaries with orbital frequencies in the mHz frequency band, allowing for detection with upcoming space-based gravitational wave detectors like \textit{TianQuin} \citep[][]{TianQinProposal2016} and the \textit{Laser Interferometer Space Antenna} \citep[LISA,][]{LISA, AmaroSeoane2022LISAastrophysics}, with DWDs acting as dominant foreground sources \citep[][]{Nissanke2012GWemissionCompactBinaries,Korol2017ProspectsDWDs, Lamberts2019DWDmodellingGW}.

Although the Galactic population of DWD binaries is expected to be in the hundreds of millions \citep[e.g.][]{Nelemans2001PopulationCloseDetachedDWDs, Marsh2011DWDsLISA, Korol2022}, only a small fraction of the observable population has been conclusively identified to date. This is largely the result of observational biases from the intrinsic dimness of WDs and from a detection preference with eclipsing DWD systems, which reveal the clear presence of a companion but require near edge-on inclinations. Ongoing and recent attempts to exploit all-sky variability surveys to find DWD and other compact binaries has resulted in an acceleration in the detection of such systems \citep[e.g.][]{BurdgeSystematic2020,VanRoestel2021AMCVnZTF,VanRoestel2022EclipsingAMCVnZTF,Kosakowski2022ZTFfourNewEclipsingWDs, Keller2022EclipsingWDbinariesGaiaZTF, Ren2023closeWDbinariesZTFGaia}. Furthermore, the extremely low mass (ELM) WD survey \citep[][]{Brown2022elmSurvey} and the supernovae type Ia progenitor survey \citep[][]{Napiwotzki2020ESOspy} have provided population statistics on the DWD sample from the DWDs discovered in their surveys. In particular, \citet[][]{Napiwotzki2020ESOspy} reveal a DWD fraction of 6\% compared to their observed WD sample, which is consistent with the fraction expected in from population synthesis \citep{Toonen2017binarityOfLocalWD}. A couple of dozen other radial-velocity variable or eclipsing DWD systems have been discovered separately to these surveys \citep[see e.g.][and references therein]{Kilic2010BinaryWhiteDwarfsthatwillMergeWithin500Myr,Hallakoun2016SDSSJ1152, Kilic2021DWDPlainSite} which typically have orbital periods between a few hours and a day. However, the most compact of DWD binaries with an orbital period less than one hour are rare. \citet{Brown2022elmSurvey} find that $\approx10\%$ of low-mass WD binaries alone have sub-hour orbital periods in the ELM WD survey. Approximately 20 out of a total of 200 DWDs have been found to be this compact, an excessive fraction compared to the $\approx5\%$ predicted \citep{Nelemans2001PopulationCloseDetachedDWDs} that is likely consistent when considering observational biases. 

Eclipsing DWD binaries offer the most precise system characterisation and those with sub-hour orbital periods have the potential for their orbital decay to be measured observationally within decades. In this paper, we present the discovery of such a compact and eclipsing DWD binary with orbital period of 47.19\,min. Its binarity was discovered using the Transiting Exoplanet Survey Satellite \citep[TESS,][]{TESS2015}, making it one of the shortest-period detached binary discovered by \textit{TESS} to date. Compared to the characterised eclipsing DWD binary sample, WDJ\,022558.21$-$692025.38 (hereafter J0225$-$6920) is located relatively nearby in the Galaxy at a distance of approximately 400\,pc; at the time of writing, J0225$-$6920 becomes the third-closest eclipsing DWD discovered after NLTT~11748 \citep[][]{Steinfadt2010NLTT11748} and GALEX~J1717+6757 \citep[][]{Vennes2011GALEXJ1717+6757}, both being of distance $\approx180$\,pc. 

Our observations are discussed in \cref{sec:observations} followed by spectral modelling in \cref{sec:Spectroscopy}. Then, we address light curve modelling and derived system parameters in \cref{sec:BinaryModelling}. We close by discussing the projected orbital decay of J0225$-$6920 in \cref{sec:gravitational_wave_source}.

\section{Observations}
\label{sec:observations}
\begin{table}
    \centering
    \caption{Observed parameters of J0225$-$6920 taken from \textit{Gaia} DR3 with reference epoch 2016-01-01. The quoted distance measurement is taken from the method described in \citet{BailerJonesDistance}.}
    \begin{tabular}{l c}
    \hline
    Parameter & Constraint\\
    \hline
        RA & 36.492414189\,$\text{deg}\pm0.035$\,mas\\
        Dec & $-69.34049573\,\text{deg}\pm0.035$\,mas\\
        PM RA & $-11.720\pm0.046$\,mas\,yr$^{-1}$\\
        PM Dec & $-25.500\pm0.049$\,mas\,yr$^{-1}$\\
        Parallax & $2.478\pm0.038$\,mas\\
        BP & $16.305\pm0.004$\,mag\\
        G & $16.398\pm0.001$\,mag\\
        RP & $16.600\pm0.007$\,mag\\
        Distance & $402.6^{+5.9}_{-6.8}$\,pc\\
        Source ID & 4693541467955966848 \\
        \hline
    \end{tabular}
    \label{tab:TableObservationsRADec}
\end{table}
\subsection{Photometry}
\label{subsec:photometry}
J0225$-$6920 was identified as a DA white dwarf by \citet[][EC02251-6933]{Kilkenny2015} and classified by \citet{GentilleFusillo2019DR2,GentilleFusillo2021WDsEDR3} as a high-probability white dwarf ($P_{\rm WD}>0.95$). J0225$-$6920 was part of a 2\,min cadence program with \textit{TESS} in Sectors 27, 28, and 29 (program G03124, TIC 631238061). During this program, we saw clear eclipsing signatures in J0225$-$6920 that showed the system to be a DWD binary, although its orbital period was not immediately clear from boxed-least-squares periodogram aliases of 23.5\,min and 47\,min. To distinguish the orbital period, we initially obtained (white light) data from the 0.41\,m PROMPT-1 telescope \citep[][]{PROMPT2005}, data from the 1.6\,m Pico dos Dias Observatory (BG40 filter) and data from the South African Astronomical Observatory (SAAO) 1.0\,m Lesedi telescope (white light). These revealed distinct primary and secondary eclipses for a 47\,min orbital period.

Following identification, we then observed J0225$-$6920 with the high-speed photometer ULTRACAM \citep{ULTRACAM2007} mounted on the ESO 3.5\,m New Technology Telescope (NTT) during five nights in July 2021. Observations were simultaneously taken using the Super SDSS $u_s$, $g_s$ and $i_s$ filters \citep{Hipercam2021}. We later observed with ULTRACAM on the NTT over three nights in September 2022 in the Super SDSS $u_s$, $g_s$ and $r_s$ filters. In total, we observed more than 15\,hrs on target with ULTRACAM. A full observing log is supplied in Appendix~\ref{tableappendix:ObservingLog}.

All data were debiased and flat-fielded; the SAAO and PROMPT-1 data with custom scripts, and the ULTRACAM data using the HiPERCAM reduction pipeline \citep{Hipercam2021}. Dark-current subtraction was additionally performed for the ULTRACAM data because of its hotter operational temperature. The flux of J0225$-$6920 was extracted through differential aperture photometry using a non-variable comparison star with \textit{Gaia} DR3 source ID 4693540643322249216. We used a variable aperture size that reflected the seeing at the time of observation set to $1.8\times$ the full-width at half-maximum of the stellar profile. All mid-exposure time-stamps were placed on a Barycentric Modified Julian Date (BMJD) Barycentric Dynamical Time (TDB) time frame.

\subsection{Spectroscopy}
\label{subsec:ObsSpectroscopy}
\begin{figure}
    \centering
    \includegraphics[trim={0cm 0.cm 0.5cm 1.5cm},clip,width=\columnwidth,keepaspectratio]{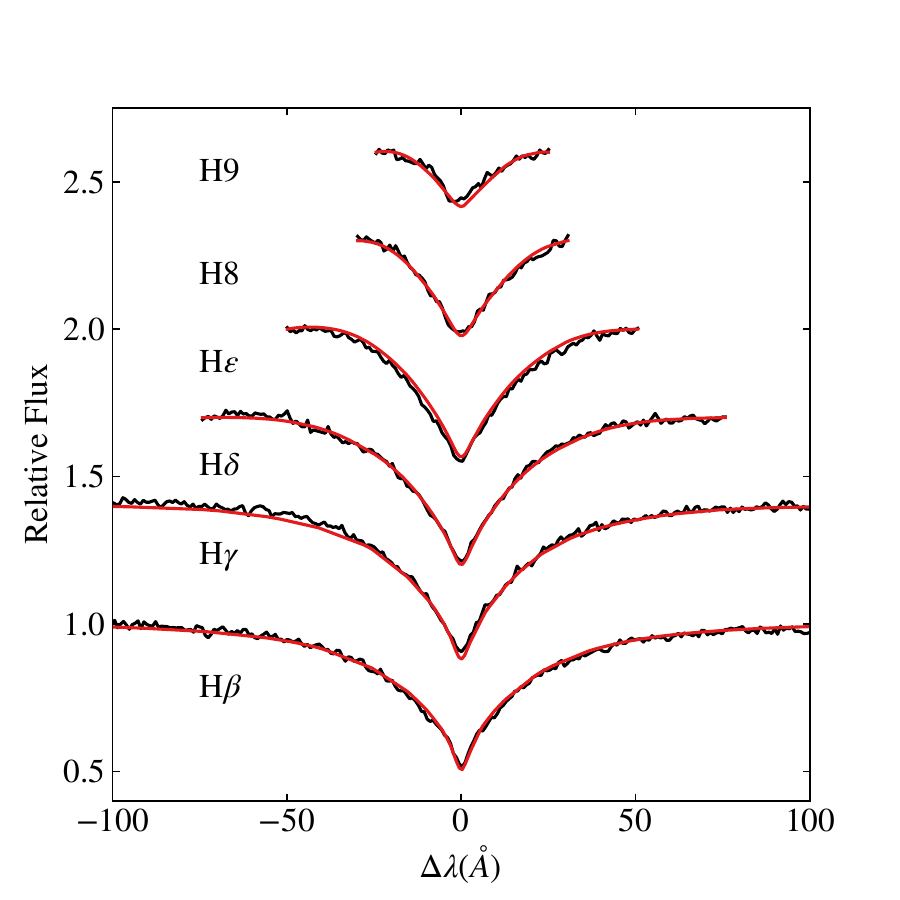}
    \caption{Normalised Balmer absorption line profiles from the combined SOAR/Goodman spectra, H$\beta$ through H$_9$. The observations are displayed in black, with the best-fit model overplotted in red. Each profile is vertically offset for clarity. The displayed model has a primary star of $T_\text{eff,1}=25\,330$\,K and $\log g_1=6.99$\,dex combined with a secondary star of $T_\text{eff,2}=13\,750$\,K and $\log g_2=7.60$\,dex. The secondary star's effective temperature that is included here was derived from a single iteration of light curve fitting using the $T_\text{eff,1}$ of a single-star fit to the co-added spectrum.}
    \label{fig:SpectrumFit}
\end{figure}

We obtained time-series long-slit spectra from the 4.1\,m SOAR telescope with the SOAR Goodman spectrograph \citep[][]{Clemens2004SOARGoodman} in October 2020. We observed with a 930 1\,mm$^{-1}$ grating and a slit-width of 1\arcsec, resulting in a resolution of $\lambda/\Delta\lambda\approx1800$ with a wavelength coverage of $3600\,\AA$--$5275\,$\AA. 43 consecutive spectra were obtained with exposure time 180\,s and a 7\,s readout time between adjacent exposures. FeAr arc lamp spectra were taken immediately before the first and immediately after the final spectrum. To avoid any possible drift to the wavelength solution extracted from FeAr arc lamps, we placed a second star (Gaia DR3 4693540643322249216) on the slit and calibrated the wavelength solution of each science exposure using the Balmer series of this comparison, i.e. the initial wavelength solution was obtained in the reference frame of the comparison star. The systemic velocity of the binary was measured from each set of calibrated spectra (see~\cref{subsec:RVs}) and the systemic velocity found through the FeAr lamp wavelength solution was used to convert from the reference frame of the comparison star to that of the barycentre, retaining the absolute wavelength calibration.

The signal-to-noise ratio (SNR) in each spectrum at the centre and wings of H$\beta$ is approximately 15 and 23, respectively. These reveal clear Balmer absorption features indicating that the brighter component is a DA~WD (see Fig.~\ref{fig:SpectrumFit}). Double-lined features in the spectrum were searched for, particularly at H$\beta$, but were not obviously apparent. No metals nor helium spectral lines were detected in the spectra.

\section{Spectroscopic analysis}
\label{sec:Spectroscopy}
\subsection{Atmospheric Parameters}
\label{subsec:atmosphericParameters}
The reduced spectra were co-added with a common rest wavelength. The normalised Balmer line profiles of the co-added spectrum were then fit with the DA~WD models of \citet[][]{Tremblay2011,Tremblay2015DA3d}. Fitting the spectrum with a single DA~WD model to represent the primary star\footnote{We follow the convention in this paper that the primary star is the hotter and brighter WD.}, we constrain the atmosphere of the brighter component\footnote{We present errors on the atmospheric constraints of the stars by combining in quadrature the statistical error from $\chi^2$ fitting with an external error of 1.2\% for $T_{\text{eff},1}$ and 0.038\,dex for $\log g_1$ \citep[][]{Liebert2005}. This is primarily to account for the error induced from flux calibration of spectra.} to have a surface gravity $\log g_1 = 7.07\pm0.04$\,dex and $T_{\text{eff},1}=24\,250\pm310$\,K.

In an attempt to retrieve the surface gravity and temperature of the secondary star, and improve the accuracy of the primary star parameters, we utilised photometric flux measurements from wide-field surveys and fit solutions for two unique WDs to the combined SED. At the time of writing, \textit{Gaia} DR3 and \textit{SkyMAPPER} \citep{Christian2018SkyMAPPER} survey data are available, which we utilise to model the photometric and spectroscopic data simultaneously. We dereddened these flux measurements with an extinction constant of $A_V=0.09$ \citep[][]{GentilleFusillo2021WDsEDR3} following the reddening prescription of \citet[][]{FitzpatrickMassa2007Extinction} with $R_V=3.1$. The WD models require a conversion from an Eddington flux to a flux observed at Earth, so we enforced a posterior distribution on the fitted distance based on the \textit{Gaia}~DR3 parallax stated in Table~\ref{tab:TableObservationsRADec}. A single DA~WD model to the dereddened \textit{Gaia} photometric fluxes yields $T_{\text{eff}}=22\,500\pm910$\,K \citep{GentilleFusillo2021WDsEDR3}, which places a minimum temperature on the primary star since the flux contributed from the cooler companion weights a single star fit to cooler temperatures. 

We then simultaneously fit all dereddened photometric flux measurements and the normalised Balmer profiles with a DA+DA model and enforced $T_{\text{eff},2}=13\,750\,$K. This temperature measurement was guided by the results of our light curve fitting in \cref{sec:BinaryModelling} when using the measured $T_{\text{eff},1}$ from a single star model to the co-added spectrum. The goals from fitting a DA+DA model were to improve the accuracy of the primary temperature by including the flux from a WD companion and to gain a measurement of the secondary's surface gravity. We obtain results of $T_{\text{eff}, 1} = 25\,330\pm330$\,K, with $\log g_1 = 6.99\pm0.04$\,dex and $\log g_2 = 7.60\pm0.23$\,dex (see Fig.~\ref{fig:SpectrumFit}). Interpolation of these parameters with WD atmosphere models indicate a primary mass of $M_1=0.35\pm0.01\,\text{M}_\odot$ for a helium core \citep{Althaus2013models} and $M_1=0.27\pm0.01\,\text{M}_\odot$ for a carbon-oxygen core \citep{Bedard2020COWDmtr} WD. A hybrid carbon-oxygen/helium core composition is also possible, and predicted masses for this composition are similar to pure-helium masses at the primary temperature \citep{Zenati2019COHeHybrid}. Although both a carbon-oxygen and helium core WD would be possible core compositions for the primary depending on the evolutionary path of the system, carbon-oxygen-core WDs are not expected below roughly 0.33\,$\text{M}_\odot$ \citep{Prada2009}, so we only consider solutions with a helium-core primary going forward. The core composition of the secondary is unclear from spectral analysis alone due to its large surface gravity error.

Lastly for atmospheric analysis, a DA+DBA WD spectral fit was performed with $T_{\text{eff},2}$ again fixed at 13\,750\,K and we found two inconsistencies. This model significantly under-predicted the flux of the weakest Balmer transitions and we found that absorption of He~\Romannum{1} (4471\,\AA) would be apparent, which it is not in the observed spectrum. We consequently predict that the companion has a hydrogen rich atmosphere, perhaps of type DA also. This DA+DA~WD prediction leads us to assume the spectroscopic results from the DA+DA model fit for the remainder of the paper, as listed in Table~\ref{tab:SED_Balmer}.

\subsection{Radial Velocities}
\label{subsec:RVs}
\begin{figure*}
    \centering
    \includegraphics[trim={1cm 0.75cm 2.25cm 11cm},clip,width=\textwidth]{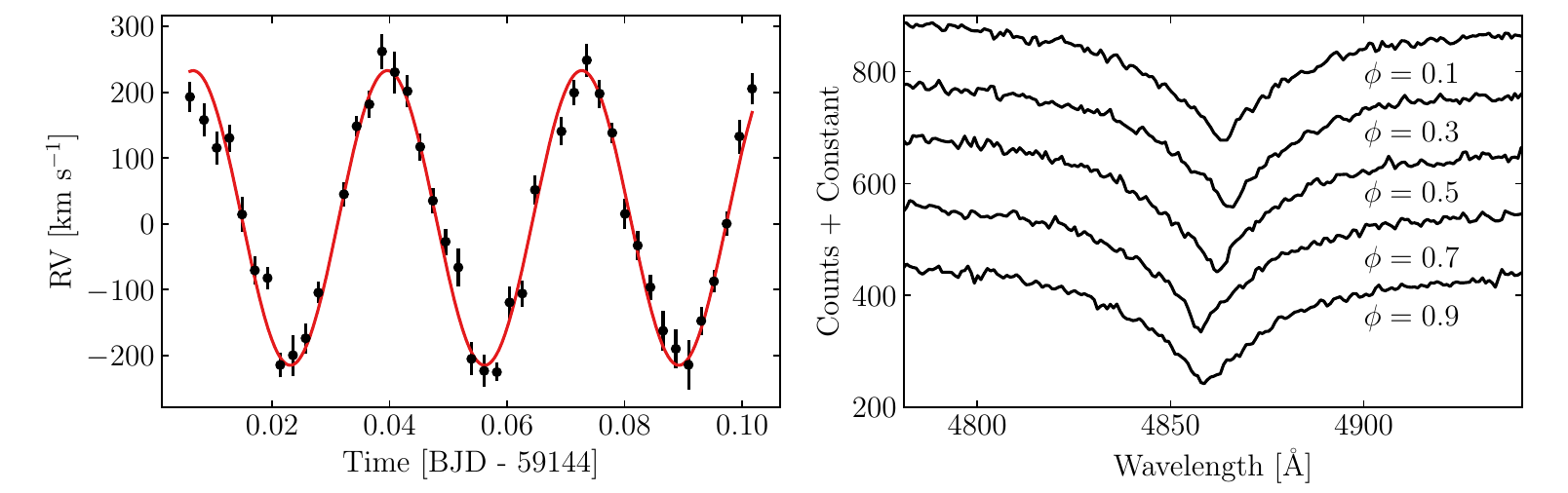}
    \caption{Left: Extracted radial velocity measurements of J0225$-$6920 with $K_1=224.0$\,km\,s$^{-1}$ and $\gamma=1.6$\,km\,s$^{-1}$ overplot. Right: Stacked observations of J0225$-$6920 as a function of phase with bin widths of 0.2 cycles, magnified on H$\beta$. A trailed spectrogram is additionally provided in Appendix~\ref{appendix:trailedSpec}.}
    \label{fig:spec_rvs_hbeta}
\end{figure*}
We then searched for any radial velocity variations in the normalised H$\beta$ absorption-line profile by modelling the full set of spectra simultaneously. The velocity of each star varies as a function of phase given by
\begin{equation}
    V_{1,2}(\phi)=\gamma \pm K_{1,2}\sin{2\pi\phi}
    \label{eqn:Velocity_fn_phase}
\end{equation}
where subscripts 1 and 2 represent the primary and the secondary stars, $\gamma$ is the systemic velocity, $K$ is the radial velocity semi-amplitude and $\phi$ is the orbital phase of the binary. The orbit is assumed circularised \citep[e.g.][]{Peters1964GravRadTwoPointMasses} and $\phi$ is solved with the centre time of the exposure and our orbital ephemeris (equation~\ref{eqn:ephemeris}). Furthermore, the line profiles themselves were fit with a 2-Gaussian component similar to all spectra\footnote{A negligible difference was found utilising a Gaussian, Lorentzian or Voigt line profile, or a combination of each to the measured radial velocities.}, each having a standard deviation, $\sigma_{\text{A,B}}$, an amplitude, $A_\text{A,B}$ and a common central wavelength solved through equation \ref{eqn:Velocity_fn_phase}. One of these Gaussian components serves to model the broad local thermal equilibrium (LTE) component of H$\beta$, while the other models the non-LTE component which is well described by a Gaussian profile. Overall, this meant that $\gamma$, $K$, $\sigma_\text{A,B}$ and $A_\text{A,B}$ were all free parameters. To reach a solution, we iteratively trialed parameter combinations using a Markov Chain Monte Carlo (MCMC) algorithm with the Python package \textsc{emcee} \citep{emcee2013}, minimising the $\chi^2$ between the model and our set of observed spectra. We chose this method to reduce the impact of the input absorption line model on the resultant measured semi-amplitude, where instead the shape of the Gaussian components are dynamically solved to when obtaining a best fit solution.

We measure $K_1=224.0\pm4.4$\,km\,s$^{-1}$ and $\gamma=1.6\pm4.1$\,km\,s$^{-1}$. For clarity, we extract radial velocity measurements independently of orbital phase (as listed in Appendix~\ref{app:RadialVelocities}) by keeping $\sigma_{A,B}$ and $A_{A,B}$ fixed from a best-fit to the stacked spectrum. Radial velocity errors are reported as the standard deviation of 1000 bootstrapping iterations. These measurements and our fit $K_1$ and $\gamma$ are plotted in Fig.~\ref{fig:spec_rvs_hbeta}. It is important to note, however, that the measured radial velocity amplitude may be underestimated due to the contribution of light from the companion in anti-phase. Even though absorption lines from the companion are not obviously apparently (see Fig.~\ref{fig:spec_rvs_hbeta}), its contribution can be significant \citep[e.g.][]{Hallakoun2016SDSSJ1152}. To check if this has a significant impact in J0225$-$6920, we searched for a consistent radial velocity amplitude across all Balmer lines by measuring those of H$\beta-\text{H}\zeta$, finding them to be within 2$\sigma$ of each other; H$\beta$ being the largest. Light curve modelling to the photometry alone agreed best with the radial velocity at H$\beta$ (see \cref{subsec:binaryModellingSetup}). As such, we report the radial velocity amplitude of H$\beta$ as that of the system.

\section{Binary Modelling}
\label{sec:BinaryModelling}
\begin{table}
    \centering
    \caption{Spectroscopically determined system parameters from the normalised Balmer line fit and the fit semi-amplitude to H$\beta$ (following the method described in \cref{sec:Spectroscopy}).}
    \begin{tabular}{l r}
    \hline
    Parameter (Spectroscopy) & Value\\
    \hline
    Primary Temperature & $T_{\textrm{eff}, 1}=25\,330\pm330$\,K\\
    Primary Surface Gravity & $\log g_1=6.99\pm0.04$\,dex\\
    Secondary Surface Gravity & $\log g_2=7.60\pm0.23$\,dex\\
    Systemic Velocity  & $\gamma=1.6\pm4.1$\,km\,s$^{-1}$\\
    Primary semi-amplitude  & $K_1=224.0\pm4.4$\,km\,s$^{-1}$\\
    \hline
    \end{tabular}
    \label{tab:SED_Balmer}
    \caption{J0225$-$6920 system parameters. All quoted measurements originate from a fit to $K_1$ and the ULTRACAM light curves only. Errors are quoted as the 16$^{\text{th}}$ and 84$^{\text{th}}$ percentiles of the post-burnin MCMC posteriors.}
    \begin{tabular}{l r}
    \hline
    Parameter (Photometry) & Value \\
    \hline
        Period & 0.03277099777(24)\,d\\
         Primary mass  & $M_1=0.40\pm0.04\,\text{M}_\odot$\\
         Primary temperature & $T_1=25\,550\pm200$\,K\\
         Primary radius  & $R_1=0.0291\pm0.0010$\,\text{R}$_\odot$\\
         Secondary mass  & $M_2=0.28\pm0.02$\,\text{M}$_\odot$\\
         Secondary temperature & $T_2=14\,350\pm100$\,K\\
         Secondary radius  & $R_2=0.0244\pm0.0007$\,R$_\odot$\\
         Inclination  & $i=85.25\pm0.06$\,deg\\
         Primary surface gravity & $\log g_1=7.11\pm0.06$\,dex\\
         Secondary surface gravity & $\log g_2 = 7.11\pm0.04$\,dex\\
         Mass ratio & $q=M_2/M_1=0.70\pm0.09$\\
         Relative squared radii & $(R_2/R_1)^2=0.70\pm0.06$\\
         Primary semi-amplitude & $K_1=240.1\pm15.6$\,km\,s$^{-1}$\\
         Secondary semi-amplitude  & $K_2=343.0\pm21.9$\,km\,s$^{-1}$\\
         \hline
    \end{tabular}
    \label{tab:Binary_params}
\end{table}

\subsection{Ephemeris}
To determine the orbital ephemeris, we simultaneously searched for an orbital period using the photometry from all detectors and filters using the multi-band Lomb-Scargle periodogram package of \citet{Vanderplas2015MultibandPhot}. By fitting individual eclipses with our best-fit synthetic light curves for the ULTRACAM data (see \cref{subsec:binaryModellingSetup}), we refined the ephemeris to a higher precision. The orbital ephemeris that we obtain, centred on the primary mid-eclipse, is
\begin{equation}
\mathrm{BMJD}_\mathrm{TDB}=59403.299199(90)  +  0.03277099777(24) E
\label{eqn:ephemeris}
\end{equation}

\subsection{PHOEBE}
\label{subsec:binaryModellingSetup}
We model the ULTRACAM light curves using the PHysics Of Eclipsing BinariEs v2.4 \citep[PHOEBE,][]{Prsa2016PHOEBE2_0, Horvat2018PHOEBE2_1, Jones2020PHOEBE2_2, Conroy2020PHOEBE2_3} package to constrain the system components. We use blackbody spectra to compute the emergent flux\footnote{WD model atmospheres are not currently available in PHOEBE. Since there is ambiguity on the secondary's contribution to the measured $T_\text{eff,1}$ (\cref{subsec:atmosphericParameters}), the spectral energy distribution for WD atmospheres would not be exact neither, making a light-curve extracted $T_\text{eff,2}$ an approximation.}. Irradiation is treated using the `Horvat' method outlined in \citet{Horvat2019Irrad} and a perfect reflection (an albedo of one) for both stars is assumed. The same Galactic reddening treatment as in \cref{subsec:atmosphericParameters} was applied, which has a small effect on the light curve morphology but is most impactful to the eclipse depths \citep[e.g.][]{Jones2020PHOEBE2_2}. The synthetic light curves do not include the contribution of gravitational lensing, however, its impact on the emergent flux is suspected to be negligible for this system, amplifying the received flux by far less than 1\% \citep[][]{Marsh2001GravitationalLensing}.

Limb darkening and gravity darkening coefficients were interpolated uniquely for the two stars using tabulated values that are specific for the Super SDSS passbands. For limb darkening, we used a power-law prescription with its coefficients set according to the 3-dimensional WD models of \citet{Claret3Dcoefficients2020andDoppler}, or the 1-dimensional WD models of \citet{Claret1Dcoefficients2020} when the 3-dimensional grid boundaries were exceeded. Gravity darkening coefficients were interpolated from the tables of \citet{Claret1Dcoefficients2020}. In each case, the interpolation was dependent on the effective temperature and the surface gravity of the respective star for a trial synthetic model.

We also reactivated the Doppler beaming functionality of PHOEBE 2.2 and pass beaming coefficients of \citet{Claret3Dcoefficients2020andDoppler}\footnote{In PHOEBE 2, this feature was disabled due to concerns of the accuracy of beaming factor computation from a non-implicit and fitted trend of the SED. In the \citet{Claret3Dcoefficients2020andDoppler} models, the beaming factor is calculated implicitly and integrated over the full SED. Therefore, the passed beaming constant is an accurate representation.}. The synthetic $K_1$ and $K_2$ are solved dynamically in PHOEBE as a function of phase (which is to say through Keplerian geometry) and incorporate the impact of gravitational redshift. The effect of smearing from a finite exposure time was corrected for assuming exposures of 9\,s in the $u_s$-band and 3\,s in the $g_s, r_s$ and $i_s$ bands (see Table \ref{tableappendix:ObservingLog}). 

To obtain a system solution, we implemented an MCMC algorithm using \textsc{EMCEE} \citep{emcee2013}, where the goodness of fit was determined by minimisation of the $\chi^2$ between the synthetic and observed light curves. We fit synthetic photometry to the ULTRACAM light curves simultaneously to improve the precision of the radii, masses and the inclination of the system. Since blackbody models are considered, the spectral energy distribution assumed for both stars in PHOEBE differs from that of a WD. The usage of blackbody models causes little issue to the $g_s$, $r_s$ and $i_s$ bands as the colour between each band for WD and blackbody spectra is small, whereas the $u_s-g_s$ colour deviates strongly. To overcome these issues, we model to the $g_s$, $r_s$ and $i_s$ bands only and we set a Gaussian prior for the primary star temperature based on our spectroscopic solution. With a final model, we then check for consistency in the $u_s$ band by allowing the temperature of the secondary to be a free parameter. The secondary temperature is identical for each of the other passbands. 

We started by modelling independently of $K_1$ and found a good fit to the photometry where the synthetic $K_1$ was within the error of that measured, giving us confidence that the measured $K_1$ in \cref{subsec:RVs} is reflective of the true value. Following this test, we introduced $K_1$ into the $\chi^2$ minimisation. Overall, the mass, temperature and radius of the primary and secondary star and the inclination of the system were free parameters in the MCMC. Only $T_{\text{eff},1}$ included a Gaussian prior while all other parameters had no set prior.

\subsection{Binary parameters}
\label{subsec:BinaryParameters}
\begin{figure*}
    \centering
    \includegraphics[trim={0cm 0cm 0 0},clip,width=\textwidth]{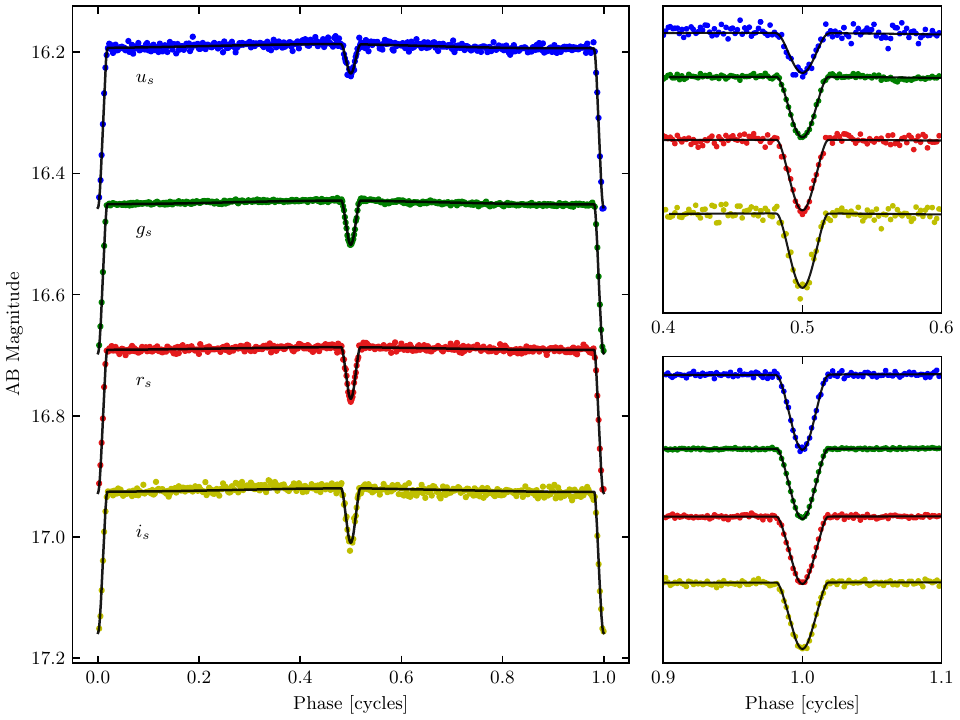}
    \caption{Left: Our ULTRACAM photometry phase-folded on the orbital ephemeris of equation~(\ref{eqn:ephemeris}) and binned for clarity. PHOEBE synthetic light curves are overlaid in black. The AB magnitude is displayed on the vertical axes where all but the $u_s$-band are vertically offset by $g_s=0.23\,\text{mag}, r_s=0.14\,\text{mag}$ and $i_s=-0.16\,\text{mag}$ for clarity. Top right/Bottom right: similar to the left-hand plot, but zoomed around the primary and secondary eclipses. Both right-hand-side plots are zoomed insets, where the light curves from each filter are manually offset in magnitude to enlarge the eclipses.}
    \label{fig:LC}
\end{figure*}
Our post-burnin results of the MCMC are presented in Table~\ref{tab:Binary_params}, with our best-fitting binary model displayed in Fig.~\ref{fig:LC}. A corner plot diagram is included in Appendix \ref{appfig:corner_LCs} showing the covariance between free parameters. A good model fit to the light curve is obtained about the eclipses and for $K_1$. Furthermore, the impact of Doppler beaming in quadrature (which causes the observed flux to be larger/smaller as a star moves towards/away from the observer) is well fit in the $u_s$-band; the band impacted most significantly \citep[see e.g.][]{Claret3Dcoefficients2020andDoppler}.

From the light curve modelling, the primary star has an effective temperature of $25\,550\pm200$\,K and for the secondary $14\,350\pm100$\,K. Component masses and radii indicate that both stars likely have helium cores, falling closely on the evolutionary tracks of helium-core WD models \citep[][]{Althaus2013models,Istrate2016}, and that the system is detached. The primary star's Roche lobe is 12\% filled and the secondary's is 18\% filled. Interpolation of the masses and the temperatures of each WD with \citet{Istrate2016} cooling track models that feature rotational and diffusional physics indicate cooling ages for each star of $t_1\approx7\,$Myr and $t_2\approx22\,$Myr, such that the more massive primary star is the younger of the two. The primary mass determined from light curve fitting is consistent with the spectroscopically predicted mass of $M_1=0.35\pm0.01\,\text{M}_\odot$ for a WD with a helium core (see \cref{subsec:atmosphericParameters}). While the $\log g_2 = 7.60\pm0.23$\,dex for a two-star spectral fit shows a worse agreement to the light-curve inferred parameters, the measurements agree at approximately a 2$\sigma$ level, where such a difference could have been influenced by the systematical errors such as smearing of the secondary star to the co-added spectrum. The modelled parameters indicate that the secondary is also likely a helium core WD.

\subsection{Similarities with known systems}
\label{sec:Discussion}
J0225$-$6920 bears a resemblance to the systems ZTF~J1749+0924, ZTF~J2029+1534 and ZTF~J0722-1839 \citep[][]{BurdgeSystematic2020}. These systems are also detached DWD binaries with similar effective temperatures and masses to J0225$-$6920, each having a shorter orbital period. All of these systems likely exited a common-envelope phase under similar conditions and follow a similar evolutionary history. In each binary, both stars are likely helium-core WDs. The characterisation of these binaries has also been used in the study of \citet[][]{ScherbakFuller2023CommonEnvelopeIsOneThirdDWD} to investigate the common envelope efficiency parameter, $\alpha$, from a second mass transfer phase. Given the similarity of J0225$-$6920, our measurements are consistent with $\alpha\approx0.2-0.4$, as inferred by these authors.

A further two compact systems that are comparable to J0225$-$6920, having similar star masses and temperatures, are SDSS~J1152+0248 \citep[][]{Hallakoun2016SDSSJ1152, Parsons2020PulsatingWDeclipsingBinary} and CSS~41177 \citep[][]{Bours2014LCmodellingCSS41177,Bours2015HST_CSS41177}. These eclipsing DWD binaries have an orbital period of 144\,min, and 167\,min, respectively. J0225$-$6920 appears to bridge the period gap between all of these DWD binaries, which can be utilised in the future to study the impact of tides on orbital decay \citep[e.g.][]{Piro2019}.

Lastly, J0225$-$6920 is much alike SDSS~J232230.20+050942.06 in regards to the core composition, being the first double helium-core, DWD, LISA verification binary discovered \citep[][]{Brown2020firstHeCoreLISAverification}. Systems such as these are vital gravitational wave sources for data quality verification of LISA when launched, given that the gravitational wave frequency and amplitude are solvable from the orbital parameters of the binary alone.

\section{J0225$-$6920 as a gravitational wave source}
\label{sec:gravitational_wave_source}
The orbit of J0225$-$6920 will decay due to a loss of orbital angular momentum from gravitational wave radiation and tides. If one assumes that the decay is dominated by gravitational wave radiation, the inspiral will obey
\begin{equation}
    \centering
    \dot f_{GW} = \frac{96}{5}~\pi^{8/3}~\left( \frac{G\mathcal{M}}{c^3}\right)^{5/3}~f_{GW}^{11/3}
    \label{eqn:fdotGR}
\end{equation}
where $f_{GW}$ is the gravitational-wave frequency, equal to twice the orbital frequency, and the chirp mass is $\mathcal{M}=(M_1M_2)^{3/5}/(M_1+M_2)^{1/5}$. Solving for our derived binary constraints in \cref{sec:BinaryModelling}, this means that the predicted orbital frequency derivative is $\dot f=\left(1.03\pm0.15\right)\times10^{-19}$\,Hz\,s$^{-1}$, or, $\dot P=\left(-8.27\pm1.21\right)\times10^{-13}$\,s\,s$^{-1}$.

The orbital decay can be precisely measured with continued observations using the time of arrival of a given orbital phase, where a number of WD binaries have been characterised in this way through continued observations \citep[][]{Barros2007HMCnc,JJ2012j0651,deMiguel2018ESCetiPdot,Burdge20197min,Burdge201920min,Burdge20208p8min,MundayHMCnc2023}. The deep eclipses of J0225$-$6920 make mid-eclipse timing the natural way to do so, where an earlier eclipse arrival time is explainable by an orbital decay. The deviation of the eclipse timing from a constant orbital period is dominated (to first order) by $\dot P t^2/2P$ for general relativistic effects, with $t$ the length of time following the first eclipse. From the data presented in this paper, the arrival time difference from the full span of data is currently a mere second, however, the bright $\textit{Gaia}$ magnitude $\left(G=16.4\,\text{mag}\right)$ and its $0.2\,\text{mag}$ deep primary eclipses makes J0225$-$6920 an exemplar system for a measured $\dot P$ that can be utilised to probe non-gravitational wave induced orbital decay in the future. This may arise in J0225$-$6920 due to tidal dissipation \citep[][]{Benacquista2011,Piro2011,FullerAndDong2012DynamicalTides,Piro2019}, in which orbital energy transferred into rotational energy of the stars is the cause of a faster orbital decay. Additionally, measurement of the orbital decay constrains the chirp mass, thus restricting the combinations of primary and secondary star masses for system modelling \citep[e.g.][]{MundayHMCnc2023}.

\begin{figure}
    \centering
    \includegraphics[trim={0cm 0.5cm 0cm 0.5cm},clip,width=\columnwidth]{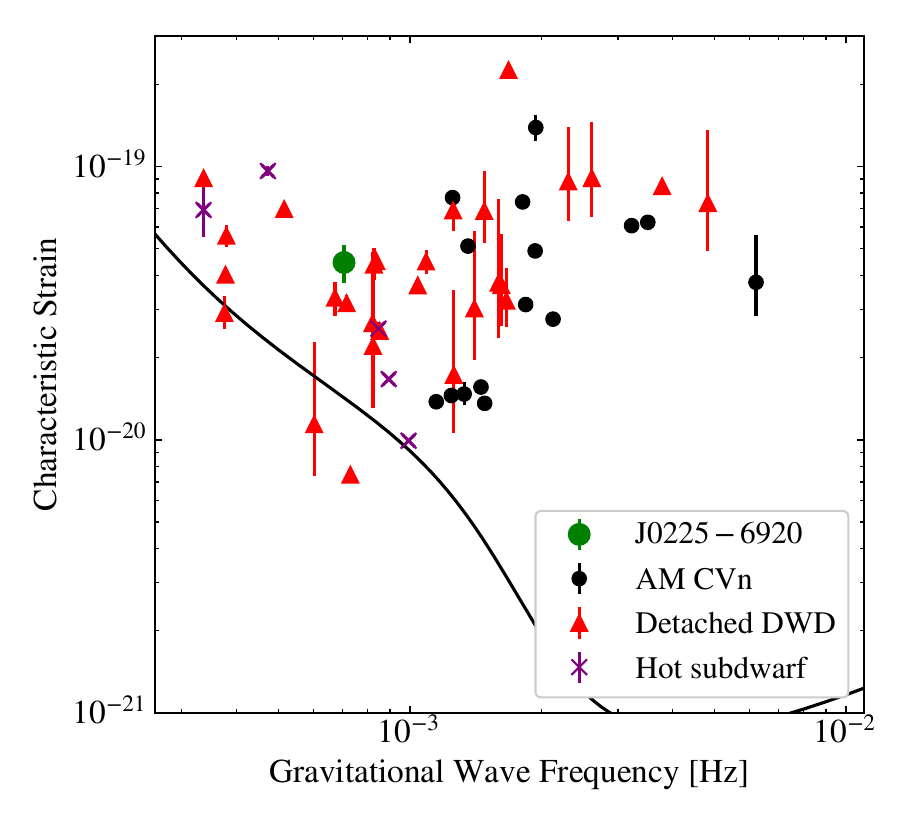}
    \caption{A plot of some of the currently known \textit{LISA} verification binary star systems, with characteristic strains predicted from a 4\,yr integration time. Systems that are listed in \citet[][see references therein]{Kupfer2023lisa} are included\protect\footnotemark[5]. These binaries are separated into the categories AM~CVn, detached DWD or binaries comprising a hot subdwarf. The addition of J0225$-$6920 to these verification binaries is depicted. The curved black line represents the predicted detection threshold of \textit{LISA} \citep{Robson2019LISAsensitivityCurve}.}
    \label{fig:LISAstrain}
\end{figure}

We find that timing of individual J0225$-$6920 primary and secondary eclipses in ideal conditions with ULTRACAM is accurate to $0.5$\,s and $2$\,s, respectively. Considering this and the predicted orbital decay, we expect to obtain a period derivative precise to better than $\approx$5\% after 5\,yrs or $\approx$1\% after 10\,yrs; the precision scaling with $t^2$. The observed deviation of the centre of eclipse arrival time will be $\approx$4\,s and $\approx$15\,s after 5\,yrs and 10\,yrs. Furthermore, the compact binary is ideally situated in the \textit{LISA} frequency band, as shown in Fig.~\ref{fig:LISAstrain}. Its distance of $402.6^{+5.9}_{-6.8}$\,pc and derived orbital parameters generates a characteristic strain of $\left(4.5\pm0.7\right)\times10^{-20}$, with a projected SNR of $1.3\pm0.2$ after a 4\,yr \textit{LISA} mission time, or $2.1\pm0.4$ after 10\,yrs \citep[][]{LEGWORK_joss,LEGWORK_apjs}. J0225$-$6920 is thus a detectable verification binary for LISA and will be the closest double helium core DWD verification binary known to date; which is a class that is expected to comprise 31\% of all LISA detectable binaries \citep{Lamberts2019DWDmodellingGW}. For comparison, the second- and third-closest double helium-core DWD verification binaries are SDSS~J063449.92+380352.2 at 435\,pc and SMSS~J033816.16$-$813929.9 at 533\,pc  \citep{Kilic2021twoLISA_zeropoint5kpc}.

\footnotetext{The data used to create Fig.~\ref{fig:LISAstrain} was obtained through \text{https://gitlab.in2p3.fr/LISA/lisa-verification-binaries}.}

In the future, J0225$-$6920 is expected to merge into a single WD, likely undergoing a hot subdwarf phase during which helium is burnt to form a carbon-oxygen core \citep{Han2002OriginsOfsdB}. For a circularised binary star system under purely general relativistic orbital decay, two stars will merge after a time \citep[][]{Peters1964GravRadTwoPointMasses}
\begin{equation}
    T_c(a_0) =\frac{5}{256} \frac{a_0^4 c^5}{G^3 M_1 M_2 (M_1+M_2)}
\end{equation}
\noindent with $a_0$ the present-day semi-major axis, $G$ the gravitational constant and $c$ the speed of light. We hence predict J0225$-$6920 to merge within $T_c(0.379\,R_\odot)=41\,$Myr.

\section{Conclusions}
\label{sec:Conclusion}
We have discovered that J0225$-$6920 is an eclipsing DWD binary with an orbital period of 47.19\,min. Multi-band light curve modelling indicates that the system likely consists of two helium-core WDs, where the primary has a DA spectral type, having a pure hydrogen surface composition, and the secondary likely does too. Its relatively close distance and brightness makes the binary a prime candidate to measure the orbital period decay in the future, key to investigate deviations from purely general relativistic orbital decay due to tides. The binary will be a detectable source for the \textit{LISA} spacecraft, and we believe it is the first such DWD verification binary discovered by the \textit{TESS} mission. J0225$-$6920 joins a small sample of fully characterised DWD binaries which will merge within a Hubble time. It is the closest double-helium-core DWD verification binary known to date; a class that is expected to account for 31\% of all LISA detectable DWDs \citep{Lamberts2019DWDmodellingGW}.

\section*{Acknowledgements}
ULTRACAM operations, SPL and VSD are funded by the Science and Technology Facilities Council (grant ST/V000853/1). JM was supported by funding from a Science and Technology Facilities Council (STFC) studentship. BB acknowledges funding through NASA award 80NSSC21K0364. IP was supported by STFC grant ST/T000406/1. SGP acknowledges the support of a STFC Ernest Rutherford Fellowship. MJG acknowledges funding by the European Research Council (ERC) under the European Union's FP7 Programme, Grant No. 833031 (PI: Dan Maoz). Support for this work was in part provided by NASA TESS Cycle 2 Grant 80NSSC20K0592 and Cycle 4 grant 80NSSC22K0737, as well as the National Science Foundation under grant No. NSF PHY-1748958. Based on observations collected at the European Southern Observatory under ESO programme 0109.D-0749(B). This paper uses observations made at the South African Astronomical Observatory (SAAO). For the purpose of open access, the authors have applied a creative commons attribution (CC BY) licence to any author accepted manuscript version arising. This research received funding from the European Research Council under the European Union’s Horizon 2020 research and innovation programme number 101002408 (MOS100PC). DJ acknowledges support from the Erasmus+ programme of the European Union under grant number 2020-1-CZ01-KA203-078200. Financial support from the National Science Centre under project No.\,UMO-2017/26/E/ST9/00703 is acknowledged.

\section*{Data Availability}
Raw time-series spectra and photometry will be made available upon reasonable request to the authors.



\bibliographystyle{mnras}
\bibliography{mnras_template} 




\clearpage
\appendix
\onecolumn
\section{Observation Log}
\begin{table*}
    \label{tableappendix:ObservingLog}
    \caption{An observing log of all J0225$-$6920 observations acquired with ULTRACAM which were subsequently used for light-curve modelling. Nights were not noticeably impacted by clouds. The duration represents the time that the telescope was on target after acquisition. MJD$_{\textrm{mid}}$ is the MJD at the centre of the observing period. In the instrument column, UCAM is an abbreviation for ULTRACAM. Observations in the $u_s$ filter were $3\times$ the cadence stated. ULTRACAM has a dead time of 24\,ms between adjacent exposures.}
    \centering
    \begin{tabular}{c c c c c c c c}
        \hline
         Night & MJD$_\textrm{mid}$  & Filters & Telescope & Instrument & Cadence (s) & Duration (min) & Comments\\
         \hline
2021-07-07  &  59403.3  &  u$_s$~g$_s$~i$_s$  &  NTT  &  UCAM   &  3.0  &  89.8  &  Seeing 1.2\arcsec  \\
2021-07-08  &  59404.3  &  u$_s$~g$_s$~i$_s$  &  NTT  &  UCAM   &  3.0  &  21.9  &  Seeing 1.3\arcsec  \\
2021-07-09  &  59405.4  &  u$_s$~g$_s$~i$_s$  &  NTT  &  UCAM   &  3.0  &  76.1  &  Seeing 1.1-1.5\arcsec  \\
2021-07-15  &  59411.4  &  u$_s$~g$_s$~i$_s$  &  NTT  &  UCAM   &  3.0  &  44.5  &  Half with 1.5\arcsec seeing, half > 2.0\arcsec \\
2021-07-16  &  59412.4  &  u$_s$~g$_s$~i$_s$  &  NTT  &  UCAM   &  3.2  &  35.6  &  Seeing 1.2\arcsec  \\
2022-09-17  &  59840.2  &  u$_s$~g$_s$~r$_s$  &  NTT  &  UCAM   &  3.3  & 316.2   &  Seeing 1.8-2.2\arcsec  \\
2022-09-18  &  59841.3  &  u$_s$~g$_s$~r$_s$  &  NTT  &  UCAM   &  3.7  &  201.6  &  Seeing 1.8-2.0\arcsec  \\
2022-09-19  &  59842.3  &  u$_s$~g$_s$~r$_s$  &  NTT  &  UCAM   &  3.2  &   142.1 &  Seeing 2.2\arcsec with short spikes  \\
         \hline
    \end{tabular}
\end{table*}


\section{Corner plot from Light Curve Fitting}
\label{appendix:corner}
\begin{figure}
\label{appfig:corner_LCs}
    \centering
    \includegraphics[keepaspectratio, width=13.9cm]{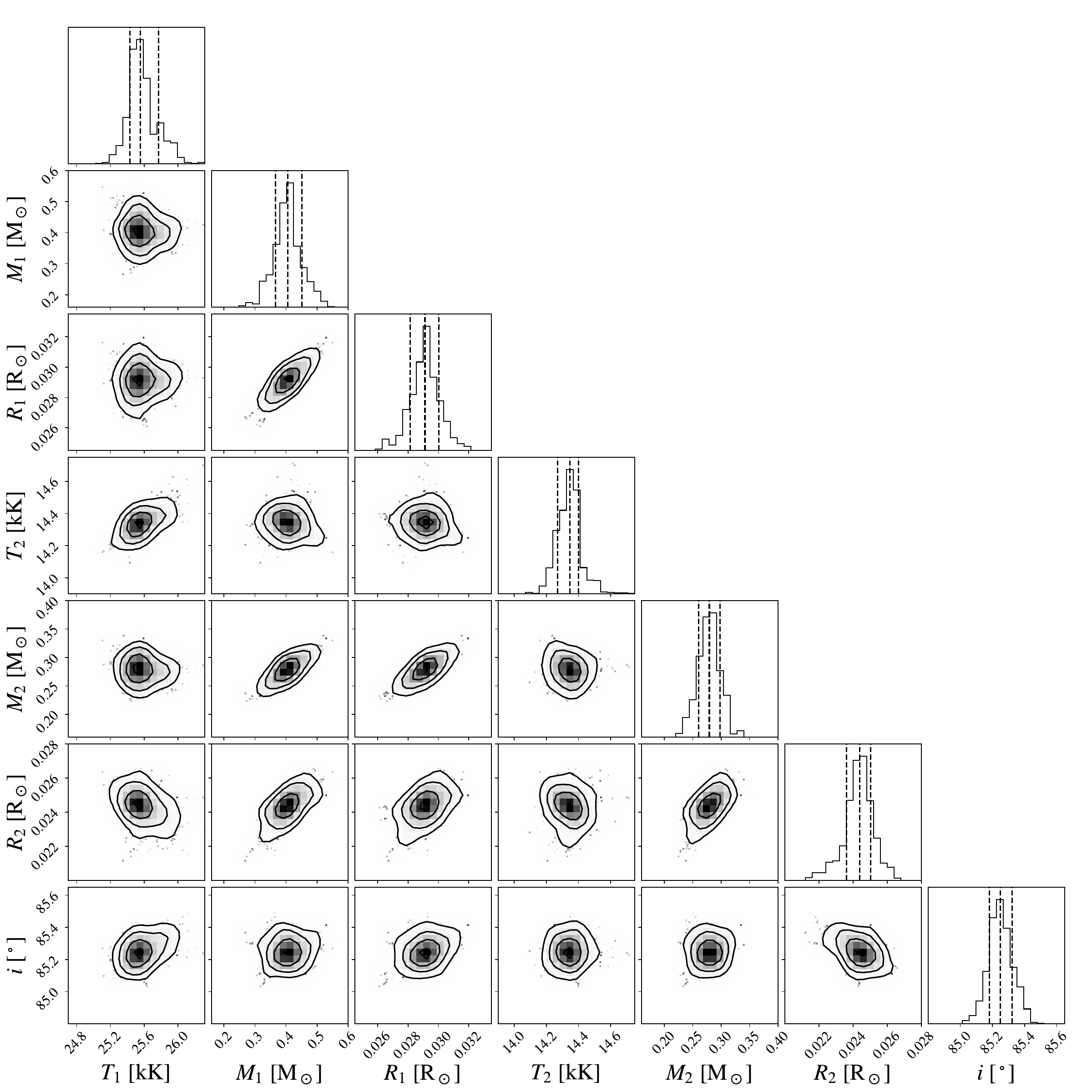}
    \caption{A corner plot diagram of our PHOEBE light-curve-modelling using an MCMC algorithm, showing the covariance of all free parameters.}
\end{figure}

\twocolumn
\section{Trailed spectrogram}
\label{appendix:trailedSpec}
\begin{figure}
\label{appfig:trailedSpec}
    \centering
    \includegraphics[keepaspectratio, width=\columnwidth]{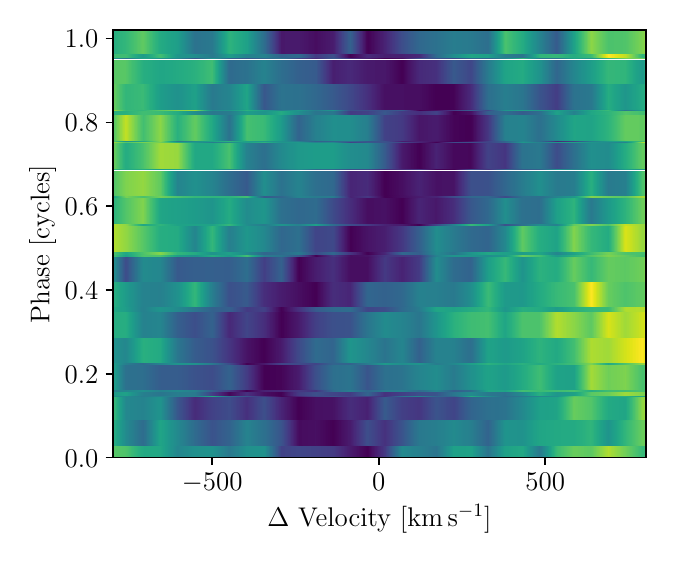}
    \caption{A trailed spectrogram of the H$\beta$ line of J0225$-$6920 from our SOAR Goodman spectra. The reference 0\,km\,s$^{-1}$ is set for a zero-velocity at wavelength 4861\,\AA.}
\end{figure}

\section{Radial Velocities}
\label{app:RadialVelocities}
\begin{table}
    \caption{The extracted radial velocity measurements of J0225$-$6920 presented in Fig.~\ref{fig:spec_rvs_hbeta}. The time-stamps presented have been corrected to a barycentric reference frame (MBJD, TDB) and the velocities (V$_\textrm{b}$) have been corrected for a barycentric velocity also.}
    \centering
    \begin{tabular}{c|c|c|c}
    \hline
        MBJD--59140 & V$_{\textrm{bary}}$ (km\,s$^{-1}$) & MBJD--59140 & V$_{\textrm{bary}}$ (km\,s$^{-1}$)\\
    \hline
4.006026 &  185.6  $\pm$  22.9  &  4.056087 &  -230.8  $\pm$  24.6\\
4.008408 &  150.5  $\pm$  25.3  &  4.05825 &  -232.7  $\pm$  14.0\\
4.010571 &  108.1  $\pm$  24.8  &  4.060413 &  -126.8  $\pm$  24.0\\
4.012734 &  123.1  $\pm$  20.8  &  4.062576 &  -113.3  $\pm$  19.7\\
4.014897 &  6.9  $\pm$  27.1  &  4.064739 &  44.2  $\pm$  21.7\\
4.01706 &  -78.1  $\pm$  22.0  &  4.069236 &  133.1  $\pm$  21.3\\
4.019222 &  -89.7  $\pm$  17.3  &  4.071399 &  192.0  $\pm$  18.9\\
4.021386 &  -221.9  $\pm$  18.9  &  4.073562 &  241.3  $\pm$  25.5\\
4.023549 &  -207.1  $\pm$  31.1  &  4.075725 &  190.5  $\pm$  21.6\\
4.025712 &  -181.3  $\pm$  23.0  &  4.077888 &  131.0  $\pm$  15.1\\
4.027875 &  -112.0  $\pm$  16.2  &  4.080051 &  7.7  $\pm$  23.5\\
4.032223 &  37.6  $\pm$  18.8  &  4.082214 &  -40.3  $\pm$  22.4\\
4.034386 &  141.0  $\pm$  15.5  &  4.084377 &  -103.7  $\pm$  19.4\\
4.036549 &  174.3  $\pm$  21.0  &  4.08654 &  -170.0  $\pm$  30.9\\
4.038712 &  254.6  $\pm$  27.1  &  4.088703 &  -197.4  $\pm$  29.2\\
4.040875 &  223.1  $\pm$  31.9  &  4.090866 &  -221.7  $\pm$  37.4\\
4.043039 &  194.2  $\pm$  24.4  &  4.093029 &  -154.9  $\pm$  21.2\\
4.045201 &  109.7  $\pm$  20.7  &  4.095192 &  -94.8  $\pm$  17.0\\
4.047364 &  27.7  $\pm$  18.8  &  4.097355 &  -7.1  $\pm$  18.8\\
4.049527 &  -34.5  $\pm$  19.9  &  4.099518 &  125.5  $\pm$  25.9\\
4.05169 &  -73.6  $\pm$  29.0  &  4.101681 &  198.0  $\pm$  24.0\\
4.053924 &  -212.7  $\pm$  25.0\\

\hline
    \end{tabular}
    \label{apptab:RadialVelocities}
\end{table}

\bsp	
\label{lastpage}
\end{document}